\documentclass[%
 reprint,
 amsmath,amssymb,
 aps,
]{revtex4-2}

\usepackage{graphicx}
\usepackage{dcolumn}
\usepackage{bm}
\usepackage{float}

\begin{document}

\preprint{APS/123-QED}

\title{Tuning Stoichiometry to Promote Formation of Binary Colloidal Superlattices}

\author{R. Allen LaCour}
\author{Timothy C. Moore}%
\author{Sharon C. Glotzer}%

 \email{sglotzer@umich.edu}
\affiliation{%
 Department of Chemical Engineering, The University of Michigan, Ann Arbor, Michigan\\
 Biointerfaces Institute, The University of Michigan, Ann Arbor, Michigan
}%

\date{\today}

\begin{abstract}
The self-assembly of binary nanoparticle superlattices from colloidal mixtures is a promising method for the fabrication of complex colloidal co-crystal structures.
However, binary mixtures often form amorphous or metastable phases instead of the thermodynamically stable phase. Here we show that in binary mixtures of differently sized spherical particles, an excess of the smaller component can promote -- and, in some cases, may be necessary for -- the self-assembly of a binary co-crystal. Using computer simulations, we identify two mechanisms responsible for this phenomenon.
First, excess small particles act like plasticizers and enable systems to reach a greater supersaturation before kinetic arrest occurs.
Second, they can disfavor competing structures that may interfere with the growth of the target structure. We find the phase behavior of simulated mixtures of hard spheres closely matches published experimental results. We demonstrate the generality of our findings for mixtures of particles of arbitrary shape by presenting a binary mixture of hard shapes that only self-assembles with an excess of the smaller component.
\end{abstract}

\maketitle


Binary colloidal mixtures are known to self-assemble into a diverse array of binary superlattices, providing a simple way to prepare colloidal co-crystals with novel combinations of properties. In many cases, including with polymer beads\cite{Bartlett1992}, microgel particles\cite{Schaertl2018}, metal nanoparticles\cite{Shevchenko2006, Boles2016b}, and quantum dots\cite{Chen2007c}, mixtures of particles differing only in their sizes can produce a compositionally ordered superlattice\cite{Murray1980, Hachisu1980, Bartlett1992, Eldridge1993a, Eldridge1995}.

The structure of the superlattice dictates important material properties, \textit{e.g.} photonic response \cite{Hynninen2007} and catalytic activity\cite{Kang2013}; thus much effort has focused on designing particles that self-assemble particular colloidal crystal structures \cite{VanAnders2015, Adorf2018InverseStructures, Pineros2018InverseAssemblies, Sherman2020InverseMaterials, Damasceno2012f, Pretti2018AssemblyParticles, Mahynski2019UsingAssembly, VanAnders2015, Adorf2018InverseStructures, Geng2019}.
However, less well understood is how to ensure that the equilibrium structure is kinetically accessible \textit{via} self-assembly.
The self-assembly of co-crystal phases appears particularly susceptible to kinetic limitations, as these phases frequently fail to assemble, instead forming glasses\cite{Kob1994ScalingMixture, RevModPhys.83.587, Dasgupta2020TuningSpheres}, or metastable phases\cite{Sanz2007EvidenceColloids, Scarlett2011AInteractions}.
Glass formation is expected when assembly kinetics are slow relative to particle mobility; metastable phases are expected when the equilibrium phase has slower assembly kinetics than thermodynamically competing phases.

Many colloidal systems are characterized by purely repulsive or hard (excluded volume) interparticle interactions, including some micron-sized colloidal spheres, polymer microgels, and nanoparticles.
Because their interactions are well characterized\cite{Bartlett1990, PhysRevE.66.060501, Royall2012b}, they are especially useful for comparing experiment with theory\cite{Eldridge1995}.
Binary mixtures of purely repulsive (hard) particles are known to resist self-assembly in many instances\cite{Coslovich2018LocalSpheres, Dasgupta2020TuningSpheres, Bommineni2020SpontaneousColloids}, but their self-assembly has been observed in experiments under certain conditions\cite{Bartlett1992, Hunt2000, Tsuneo2009, Schaertl2018}.
Understanding why self-assembly occurs in some situations but not others is necessary for further advances.

In this Letter, we demonstrate using computer simulation that variation of the stoichiometry can enhance the kinetics of co-crystal self-assembly in binary mixtures whose components differ in size.  Self-assembly of binary crystals is usually attempted ``on-stoichiometry," in which the initial fluid phase has the same stoichiometry as the target crystal\cite{Khadilkar2012a, Bommineni2017, Bommineni2020SpontaneousColloids, Dasgupta2020TuningSpheres, Coli2021AnCrystal}. We show that going ``off-stoichiometry" by adding an excess of the smaller component can dramatically improve self-assembly.
We demonstrate that this enhancement can be attributed to two mechanisms, both of which we observe in our simulations. Specifically, we show that the excess of small particles (i) enables the large component to remain mobile at higher supersaturation, facilitating self-assembly of the equilibrium structure and avoiding kinetic arrest; and (ii) can disfavor competing structures that may interfere with the growth of the equilibrium structure.

We first investigate a binary inverse power law (IPL) system at a size ratio ($\gamma$) of 0.55, similar to many experiments\cite{Hachisu1980, Bartlett1992, Hunt2000, Shevchenko2006}. 
Setting the power $n$ to 50 makes the particles similar in softness (steepness of repulsion with interparticle distance -- less steep is softer) to some experimental microgels\cite{Schaertl2018} but slightly softer than most PMMA beads\cite{Royall2012}. 
We make them slightly soft so as to be able to use standard molecular dynamics (MD) algorithms; from our previous work\cite{LaCour2019b} and the phase diagram computed here, we do not expect their phase behavior to deviate significantly from hard spheres. 
We used HOOMD-Blue\cite{Anderson2019b, Glaser2015a} to conduct, freud\cite{Ramasubramani2020Freud:Data} to analyze, and signac\cite{Adorf2018b} to organize the MD simulations. A detailed description of our simulation methodology is provided in section S1 of the supplementary material\cite{Martyna1994, Vega2007, Phillips2011, Frenkel1984, Steinhardt1983Bond-orientationalGlasses,Ramasubramani2020Freud:Data, Stukowski2010, Adorf2018b, Anderson2016a}.
We describe stoichiometry throughout this work in two ways: using the ratio $N_L$:$N_S$, where $N_L$ and $N_S$ are the number of large and small particles respectively, or using the fraction of small particles $x_S = N_S/(N_L+N_S)$.

\begin{figure}[htbp]
    \centering
    \includegraphics[scale=1.0]{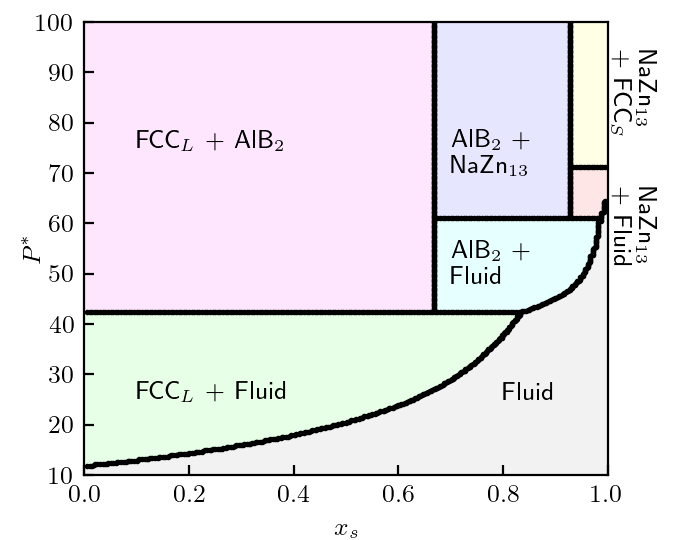}

    \caption{Thermodynamic phase diagram for the binary inverse power law model (IPL) at $\gamma = 0.55$, $n=50$, and $kT/\epsilon=1$. Five phases are present: fluid, FCC$_L$, FCC$_S$, AlB$_2$, and NaZn$_{13}$. Together they form 7 distinct regions.}
    \label{fig:Figure1}
\end{figure}

Via free energy calculations\cite{Frenkel1984, Vega2007}, we computed the thermodynamic phase diagram of the binary IPL model at $kT/\epsilon = 1$, as shown in Figure \ref{fig:Figure1}, plotted in terms of reduced pressure $P^*=P\sigma^3/\epsilon$ and $x_S$, where $\epsilon$ and $\sigma$ are the energy and length scales of the IPL.
Because of comparable experimental\cite{Bartlett1992} and simulation\cite{Eldridge1993a} studies, we considered the following candidate phases: a face-centered cubic crystal of the large particles (FCC$_L$), a face-centered cubic crystal of the small particles (FCC$_S$), an AlB$_2$ co-crystal, and a NaZn$_{13}$ co-crystal. Their stoichiometries $N_L$:$N_S$ are 1:0, 0:1, 1:2, and 1:13, respectively.

The phase diagram tells us the equilibrium phase(s) for a given set of conditions, but does not tell us whether the phases are kinetically accessible.
For self-assembly to occur, the average time for another phase to nucleate and grow must be shorter than the time accessible in experiment (or simulation). Both nucleation and growth rates are strongly influenced by the degree of supersaturation. For a fluid-to-solid transition, increasing the degree of supersaturation has contrasting effects: the free energy barrier for nucleation decreases, favoring assembly, but the particle mobility decreases, disfavoring assembly\cite{Sosso2016a}. If the mobility decreases too much before the free energy barrier becomes surmountable, the particles become kinetically arrested, inhibiting the formation of the equilibrium solid phase.

\begin{figure*}[htbp]
    \centering
    \includegraphics[scale=1.0]{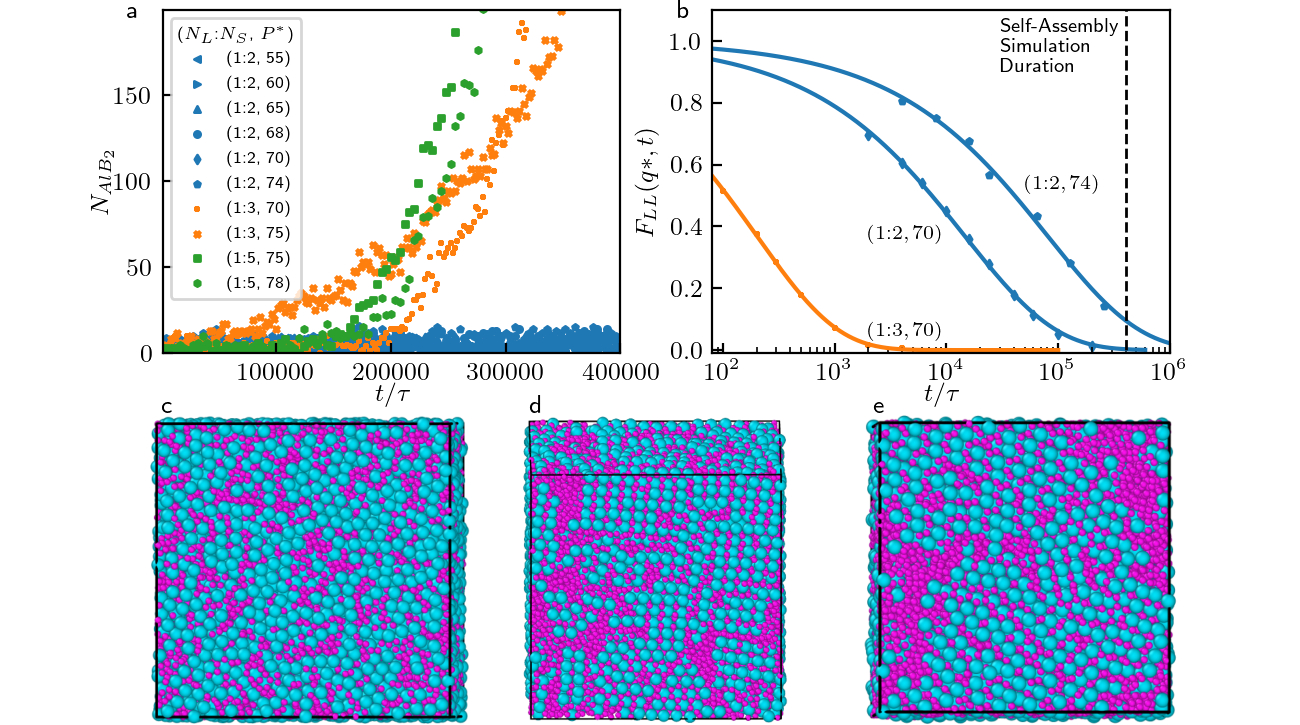}

    \caption{Self-assembly of AlB$_2$. The plot in a) shows the evolution of the number of large particles identified as AlB$_2$ for NPT simulations at the given pressure and stoichiometry.
    All simulations at $N_L$:$N_S$ = 1:2 (colored blue on the plot) overlap substantially because they never exceed $N_{AlB_2}$ = 16.
    The plot in b) shows the decay of the intermediate scattering function for certain combinations of stoichiometry and pressure. 
    The lines are fits to the data. The dotted black line indicates the duration of simulations in a).
    Snapshots of the results are shown for stoichiometries and pressures of c) $N_L$:$N_S$ = 1:2, $P^*=70$, d) $N_L$:$N_S$ = 1:3, $P^*=70$, and e) $N_L$:$N_S$ = 1:5, $P^*=75$. The simulations all began in a fluid state. 
    }
    \label{fig:Figure2}
\end{figure*}

We first investigate whether AlB$_2$ will homogeneously nucleate from a fluid for a variety of pressures and stoichiometries. 
The simulations were initialized in a fluid-like state with 27,000 particles at constant temperature $T$ and pressure $P$ (i.e., an NPT ensemble), and run for 4$\cdot10^5\tau$ timesteps, where $\tau = \sigma(m/\epsilon)^{1/2}$ and $m$ is particle mass.
Because we observed some crystal growth at $N_L$:$N_S$ = 1:3 and $P^* = 70$ and wanted to verify that the crystal continued to grow, we continued that simulation for an additional 4$\cdot10^5\tau$ timesteps.
In Figure \ref{fig:Figure2}a we show the evolution in the number of AlB$_2$-like particles up to 200 particles (according to our order parameter; see section S2 of the supplementary material) to observe the initial growth of the co-crystal nuclei. 

Figure \ref{fig:Figure2}a shows that $N_{AlB_2}$ never exceeds 16 for the on-stoichiometry systems at the chosen pressures, indicating that self-assembly never occurs. In contrast, we find that $N_{AlB_2}$ increases to 200 and beyond for systems with an excess of small particles.
The results are consistent with the system snapshots shown in Figures \ref{fig:Figure2}c-e, where crystals grains are only apparent at 1:3 and 1:5.
We note the presence of small grain sizes, which mirrors the results obtained by Bommineni et al. with particle swap moves\cite{Bommineni2019ComplexSpheres.} in binary mixtures of hard spheres.
At the highest pressure we simulated for $N_L$:$N_S$ = 1:2 ($P^* = 74$), particle mobility is extremely limited, as shown in Figure \ref{fig:Figure2}b where we plot the temporal decay of the first peak ($q*$) in the intermediate scattering function calculated for the large particles ($F_{LL}(t)$).
We thus conclude that self-assembly is only possible with an excess of small particles on the time scale of our simulations. We attribute this result in part to particles being more mobile at higher $x_s$. 
For example, by fitting the decay of $F_{LL}(t)$ to a stretched exponential (indicated by the lines in Figure \ref{fig:Figure2}b), we computed that the structural relaxation time is around 76 times longer at a stoichiometry of 1:2 than at 1:3 (14,700$\tau$ versus 202$\tau$)  at $P^*=70$, which indicates much slower equilibration at 1:2.

We next analyze the growth of AlB$_2$ in the presence of crystalline seeds. By construction, these simulations bypass the need to form a critical nucleus and thus may allow self-assembly on shorter time scales than required for homogeneous nucleation.
Each simulation was prepared by compressing a fluid around a perfect (constructed) seed of AlB$_2$ and then allowing the fluid and seed to evolve in an NPT simulation. The seed crystals were chosen to be small but post-critical, as evidenced by their persistence in the simulations.

\begin{figure*}[htbp]
    \centering
    \includegraphics[scale=1.0]{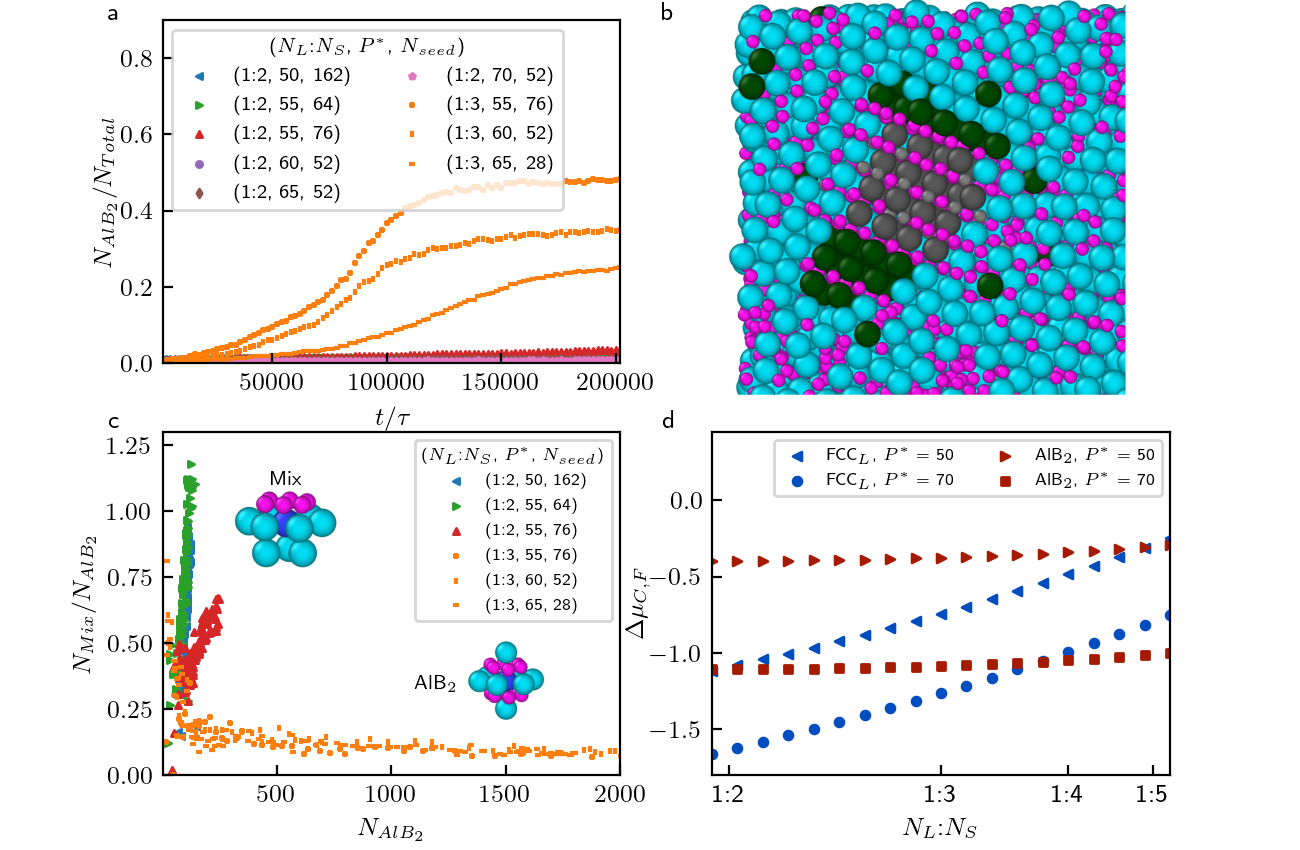}
    \caption{Crystal growth in seeded simulations. 
    The plot in a) shows the evolution of the number of large particles identified as AlB$_2$ from seeded simulations for different $x_s$, $P\sigma^3/\epsilon$, and initial seed size ($N_{seed}$).
    The image in b) is a snapshot of the end of the seeded simulation at $N_L$:$N_S$ = 1:2 and $P\sigma^3/\epsilon=55$. 
    Large and small particles belonging to the initial seed are colored dark grey and light grey, respectively; large particles classified as mixed FCC-AlB$_2$ are colored dark green.
    The plot in c) shows the number of particles classified as AlB$_2$ ($N_{AlB_2}$) versus the ratio of the number classified as mixed FCC-AlB$_2$ to $N_{AlB_2}$ ($N_{Mix}/N_{AlB_2}$).
    The insets  illustrate the mixed FCC-AlB$_2$ and AlB$_2$ environments.
    The plot in d) shows the chemical potential driving force $\Delta\mu_{C,F}$ for the FCC$_L$ and AlB$_2$ as a function of pressure and stoichiometry, where $\Delta\mu_{C,F}$ is defined by Equation \ref{eq:u}. Errors (calculated as described in S1 of the supplementary material) are smaller than the size of the points.
    }
    \label{fig:Figure3}
\end{figure*}

Figure \ref{fig:Figure3}a shows the evolution of the fraction of large particles classified as AlB$_2$-like ($N_{AlB_2}/N_{Total}$). 
We consistently find more crystal growth off-stoichiometry at $N_L$:$N_S$ = 1:3, with final values of $N_{AlB_2}/N_{Total}$ ranging from 0.25 to 0.48, than on-stoichiometry, for which $N_{AlB_2}/N_{Total}$ never rises above 0.035.

Inspection of the growing seeds at $N_L$:$N_S$ = 1:2 revealed the accumulation of non-AlB$_2$ layers of particles on the seed (an example at $P^*= 55$ is shown in Figure \ref{fig:Figure3}b).
We identified many of these layers to be two (or more) subsequent close-packed planes of large  particles.
This possibility seemed likely because FCC$_L$, which consists of close-packed planes, is metastable under the conditions we investigate, and AlB$_2$ has a close-packed layer of large particles in its structure onto which additional close-packed layers could grow.
We call a layer of these particles a ``mixed layer" and the associated coordination environment ``mixed FCC-AlB$_2$;" we denote the number of these particles $N_{mix}$.
In Figure \ref{fig:Figure3}b we illustrate their presence in dark green for a seed grown at $P^* = 55$ and $N_L$:$N_S$ = 1:2.

We quantify the formation of the mixed layer during the seeded simulations in Figure \ref{fig:Figure3}c, plotting $N_{mix}/N_{AlB_2}$ versus $N_{AlB_2}$. 
For $N_L$:$N_S$ = 1:2 we plot only the results for $P\sigma^3/\epsilon \leq 55$ because at higher pressures $N_{AlB_2}$ never exceeds 100 (\textit{i.e.}, those seeds grow negligibly over the simulation).
Off-stoichiometry at $N_L$:$N_S$ = 1:3, the proportion of mixed layers decreases with crystal growth in all cases. 
In contrast, on-stoichiometry at 1:2 the proportion always increases, indicating that mixed layers form more frequently than AlB$_2$ layers.

We identify a thermodynamic reason as to why the mixed layers are more prevalent at $N_L$:$N_S$ = 1:2.
Because the mixed layer is essentially the formation of an FCC layer where an AlB$_2$ layer should have formed, its appearance likely correlates with the thermodynamic stability of the competing FCC$_L$ phase. 
In Figure \ref{fig:Figure3}d we examine the chemical potential difference between the particles in the fluid and the solid:
\begin{equation}
    \Delta\mu_{C,F} =
    \mu_{C}
    - (1 - x_{C})\mu_{F}^L
    - x_{C}\cdot\mu_{F}^S
    \label{eq:u}
\end{equation}
The quantity $\mu_{C}$ is the chemical potential of the crystal; $x_{C}$ is the fraction of small particles in the crystal; and $\mu_{F}^L$ and $\mu_{F}^S$ are the chemical potentials of the large and small species in the fluid, respectively.
More negative $\Delta\mu_{C,F}$ values indicate stronger thermodynamic driving forces for nucleus formation.

Figure \ref{fig:Figure3}d shows that the $\Delta\mu_{C,F}$ of both crystals decreases with pressure but increases with a greater proportion of small particles.
However, we find that $\Delta\mu_{FCC_L,F}$ is more sensitive to stoichiometry than $\Delta\mu_{AlB_2,F}$.
For example, at $P^* = 70$, changing the stoichiometry from 1:2 to 1:3 increases the $\Delta\mu_{C,F}$ of FCC$_L$ by 0.37 kT while only increasing the $\Delta\mu_{C,F}$ of AlB$_2$ by 0.02 kT, 
resulting in a greater preference of the fluid to form AlB$_2$ relative to FCC$_L$.

To summarize these results, we find that AlB$_2$ does not self-assemble or even grow from a seed crystal in on-stoichiometry fluid.
We identified two reasons its formation is inhibited: slow dynamics and interference from a competing phase. Both issues are alleviated by adding excess small particles.

Our simulations should be most comparable with the experiments of Bartlett et al.\cite{Bartlett1990, Bartlett1992} using PMMA particles because our results are for a similar size ratio (0.55 vs. 0.58) and they explore how stoichiometry affects assembly.
In Table \ref{tab:stoichio}, we compare the binary crystals we obtain with theirs.
Our results at $N_L$:$N_S$ of 1:2, 1:3, 1:5, are shown in Figure \ref{fig:Figure2}; results for the other stoichiometries are shown in the section S3 of the supplementary material. 
We denote any experiment not reported with ``-".

\begin{table}
	\centering
	\caption{Crystals Observed in Simulation and Experiment}
	\label{tab:stoichio}
	\begin{tabular}{llllll}
		\hline
		\multicolumn{1}{l}{$N_L$:$N_S$} &  \multicolumn{1}{l}{Sim. Structures}
		&  \multicolumn{1}{l}{Exp. Structures$^*$}\\
		\hline
		1:2 & Amorphous & Amorphous\\
		1:3 & AlB$_2$ & -\\
		1:4 & AlB$_2$ & AlB$_2$\\
		1:5 & AlB$_2$ & -\\
		1:6 & AlB$_2$ & AlB$_2$\\
		1:9 & AlB$_2$/NaZn$_{13}$ & NaZn$_{13}$\\
		1:13 & NaZn$_{13}$ & -\\
		1:14  & NaZn$_{13}$ & NaZn$_{13}$\\ 
		1:20 & NaZn$_{13}$ & NaZn$_{13}$\\
		1:30 & NaZn$_{13}$ & NaZn$_{13}$\\
		\hline
		$^*$Bartlett et al.\cite{Bartlett1992}
	\end{tabular}
\end{table}

Overall, we see strong agreement between simulation and the published experimental results. 
We both obtain an amorphous structure at 1:2, but see AlB$_2$ with a slight excess of small particles.
Around a stoichiometry of 1:9, we both begin to see NaZn$_{13}$ self-assemble, and continue to see it self-assemble at stoichiometries up to 1:30.

\begin{figure}[htbp]
    \centering
    \includegraphics[scale=0.93]{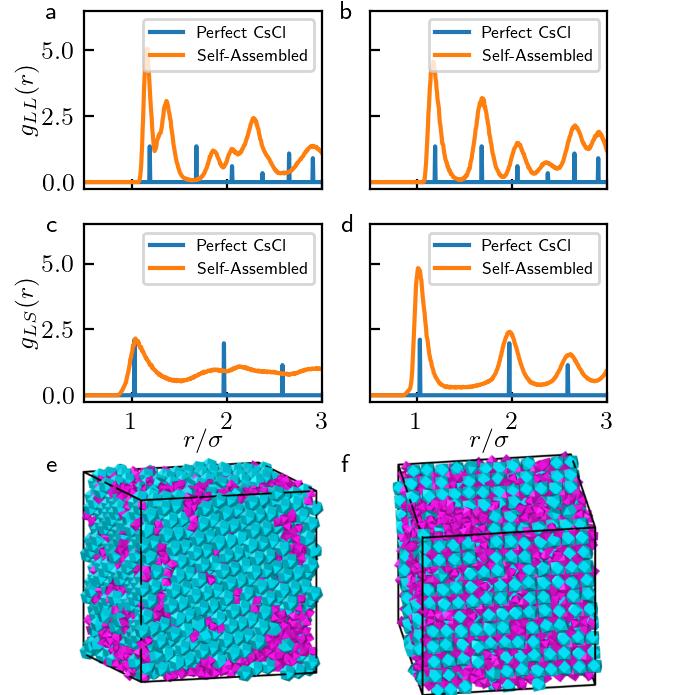}
    \caption{The self-assembly of hard cuboctahedra and octahedra. 
    The plots show the radial distribution functions (RDFs) averaged over the final few frames of self-assembly at stoichiometries of a,c) 1:1 and b,d) 1:2. 
    Also shown are the RDFs for a perfect CsCl structure. 
    We show the RDF for large particles ($g_{LL}$) and for large and small particles ($g_{LS}$); the RDF for small particles is dominated by fluid-like small particles. 
    Snapshots of the results are shown for e) 1:1 and f) 1:2. 
    Simulations were run in an NPT ensemble with 4096 particles under a slow compression starting at a volume fraction of 0.565.}
    \label{fig:Figure4}
\end{figure}

To establish whether the self-assembly of other binary crystals may be assisted by an excess of small particles, we also simulated a binary mixture of hard cuboctahedra and octahedra at a volume ratio of 5:1, which has proven difficult to crystallize despite being capable of comprising a space-filling CsCl-type structure.\cite{Khadilkar2012a}
Like the IPL spheres, the particles in this system are purely repulsive (hard).
To our knowledge, the CsCl structure has never been self-assembled from these hard shapes; in previous work, attractive patches on the particles' surface were required for self-assembly\cite{Escobedo2016EffectCompound}.
In Figure $\ref{fig:Figure4}$, we present our results for self-assembly conducted at stoichiometries of 1:1 and 1:2.
Due to the higher computational cost of simulating anisotropic particles\cite{Anderson2016a, ramasubramani2020mean}, we used a slow compression scheme with 4096 particles (details in section S1 of the supplementary material).

By comparison with the RDFs of perfect CsCl, we identified the result at 1:2 to be CsCl. 
At 1:1, a single-component structure composed of the large particles self-assembles, while the small particles remain fluid-like.
It is thus apparent that, although particle mobility is not limited, the single-component structure (successfully) competes with CsCl when the fluid is on-stoichiometry, and an excess of small particles is necessary to observe the thermodynamically preferred binary structure. 

In summary, we demonstrated that the self-assembly of binary nanoparticle superlattices can be promoted by adding an excess of the smaller component to the colloidal fluid mixture. While some crystals, like NaZn$_{13}$, do not require an excess of small particles, the surprisingly dramatic influence of excess small particles on the assembly of AlB$_2$ and CsCl suggests many other binary systems may best -- or only -- self-assemble off-stoichiometry. Our results likely apply best to purely repulsive systems; we will examine attractive systems in future work.

The software used to carry out this research was developed with support from the National Science Foundation, Division of Materials Research Award No. DMR 1808342. The investigation of co-crystallization was supported by the Department of the Navy, Office of Naval Research under ONR award number N00014-18-1-2497. 
This work used resources from the Extreme Science and Engineering Discovery Environment (XSEDE)\cite{XSEDE}, which is supported by National Science Foundation grant number ACI-1548562; XSEDE Award DMR 140129; and also used resources of the Oak Ridge Leadership Computing Facility, which is a DOE Office of Science User Facility supported under Contract DE-AC05-00OR22725. Additional computational resources supported by Advanced Research Computing at the University of Michigan. 
\bibliography{bibliography}

\end{document}


\beginsupplement

\section{Simulation Methods}

 We used molecular dynamics (MD) with the HOOMD-Blue simulation toolkit\cite{Anderson2008a, Glaser2015a} to study the binary inverse power law (IPL) potential with power $n=50$:
\begin{equation}
    U(r_{ij})=\epsilon\Big(\frac{\sigma_{ij}}{r_{ij}}\Big)^{50}.
    \label{eq:pot}
\end{equation}
The quantity $U$ is the potential energy of interaction between two particles $i$ and $j$ at a distance of $r_{ij}$. The quantity $\sigma_{ij}$ represents the diameter of the particles; for interactions between unlike particles, we set it to their average diameter. We denote the diameter of the large particles as $\sigma$ and diameter ratio between small and large particles as $\gamma$.
The unit of energy is $\epsilon$, which is set to 1 temperature unit ($kT$) throughout this work.
The IPL exhibits thermodynamic scaling between temperature and pressure such that a change in temperature can be mapped to an equivalent change in pressure\cite{Hoover1970}; thus we only need to investigate its behavior at a single temperature.
 We truncated the potential at a cutoff of 1.3$\sigma_{ij}$ and shifted it from an energy of  2.0$\cdot 10^{-6}\epsilon$  at the cutoff to zero.

 We used NPT simulations based on the MTK equations\cite{Martyna1994} to collect data on nucleation from a fluid and growth from a seed and when computing the intermediate scattering function between large particles ($F_{LL}(q,t)$).

 We used HOOMD-Blue's Hard Particle Monte Carlo module to simulate the mixtures of cuboctahedra and octahedra.
  On-stoichiometry simulations at $N_L$:$N_S$ = 1:1 consisted of a slow compression from a packing fraction of 0.565 to 0.635 in increments of 0.001.
 Off-stoichiometry simulations at $N_L$:$N_S$ = 1:2 consisted of a slow compression from a packing fraction of 0.565 to 0.615 in increments of 0.001.
 We compressed to a higher packing fraction at $N_L$:$N_S$ = 1:1 because the system had not finished crystallizing at 0.615, while the system at $N_L$:$N_S$ = 1:2 was crystalline by that packing fraction.

 Free energy calculations were performed with HOOMD-blue to obtain the phase diagram shown in Figure 2.
 Pressure-Volume (PV) data were gathered for the fluid and solid phase from NVT and NPT simulations respectively.
 Free energies at different $P\sigma^3/\epsilon$ were computed by integrating curves fit to the PV data.
 The reference free energy for the fluid was taken to be that of a dilute gas; the reference free energy for the solid was computed from the Einstein molecule method\cite{Vega2007}, a variant of the Frenkel-Ladd method\cite{Frenkel1984} in which a single particle is fixed instead of the center of mass.
By comparing free energies differences computed by PV integration to those from two Einstein molecule method calculations, we estimated our errors to be below 0.05 $kT$.
The Langevin integrator\cite{Phillips2011} within HOOMD-Blue was used when performing the Einstein molecule method.

 We used Steinhardt order parameters\cite{Steinhardt1983Bond-orientationalGlasses} to classify particles as having AlB$_2$, mixed FCC-AlB$_2$, or fluid environments.
 We discuss our specific use in section S2 of the Supplementary Material.
The freud software library\cite{Ramasubramani2020Freud:Data} was used to calculate radial distribution functions and Steinhardt order parameters.
We used Ovito\cite{Stukowski2010} to visualize particles throughout the work.

The computational workflow was supported by the signac data management framework \cite{Adorf2018b}.

\section{Order parameter}

\begin{figure}[htpb]
  \centering
  \includegraphics[scale=0.8]{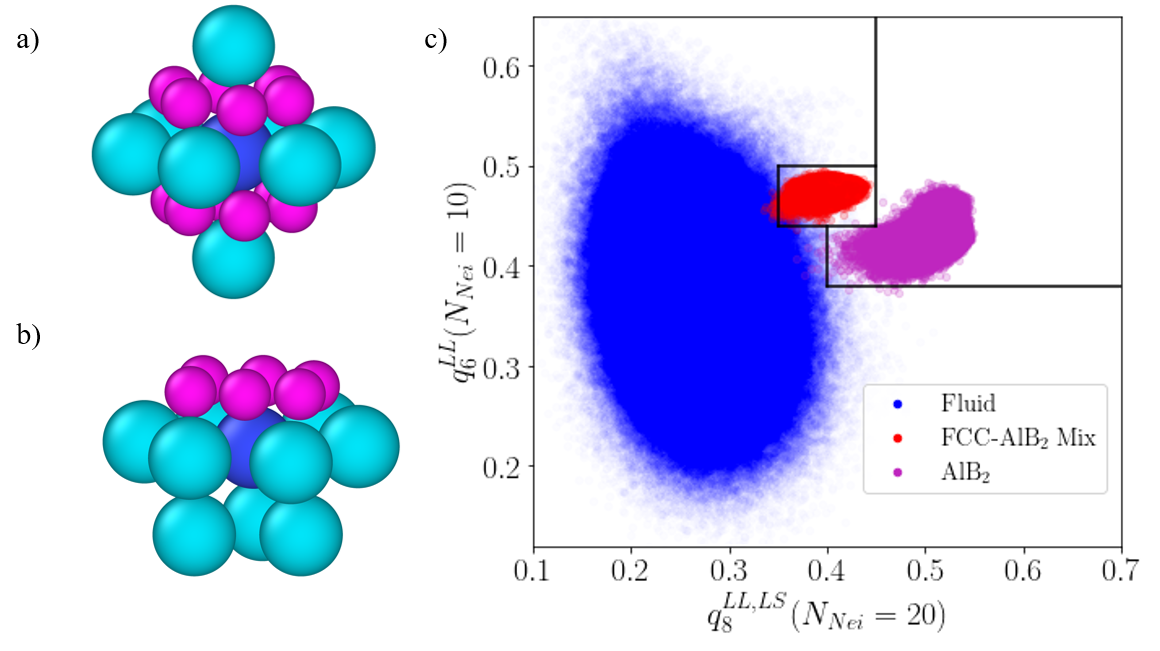}
  \caption{The coordination of large particles in a) AlB$_2$ and b) mixed FCC-AlB$_2$ environments; c) the Steinhardt order parameters of large particles in those environments and a fluid environment at $N_L$:$N_S$ = 1:2.
  In a) and b) the reference particle is colored dark blue
  We use the Steinhardt order parameter $q_8$ for the first 20 neighbors (the number of neighbors of each large particle in the perfect AlB$_2$ crystal) of either type, and the $q_6$ for the first 10 $large$ neighbors.
  The data for AlB$_2$ and mixed FCC-AlB$_2$ were generated from simulations of pre-assembled versions of the structures.
  The structure used for mixed FCC-AlB$_2$ is shown in Figure S4.
  The black lines in c correspond to how we classified particles.
  The distribution of order parameters shown in c) is computed at $P^* = 60$.}
  \label{fig:op}
\end{figure}

\begin{figure}[htbp]
    \centering
    \includegraphics[scale=0.85]{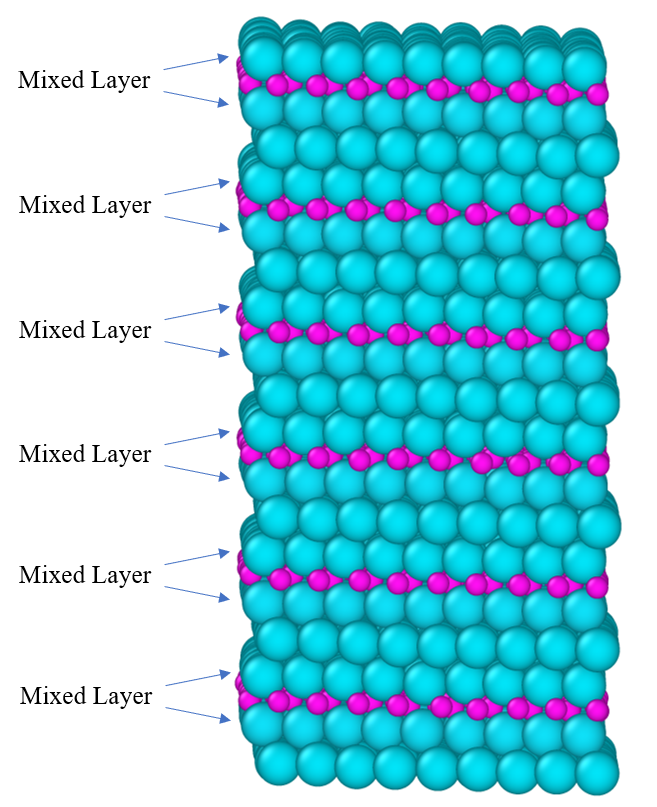}
    \caption{Snapshot of the structure we used to determine the order parameter for ``mixed FCC-AlB$_2$" environments. We only computed the order parameter for layers denote "Mixed Layer".}
    \label{fig:mixed_env}
\end{figure}

We use Steinhardt order parameters to classify particles according to their coordination environment.
We show in Figure \ref{fig:op} that a combination of two order parameters can distinguish whether large particles are in fluid, AlB$_2$ (Figure \ref{fig:op}a), or ``mixed FCC-AlB$_2$" (Figure \ref{fig:op}b) environments, which we show to be a common defect during AlB$_2$ self-assembly.
The black lines in Figure \ref{fig:op}c show how we classify particles with specific $q^{LL}_6$ and $q^{LL,LS}_8$ into the three environments.
For the mixed FCC-AlB$_2$ environments, we simulated a structure involving alternating layers of a hexagonally close packed crystal and AlB$_2$ crystals and computed the order of the particles at the interface; a visualization of the structure is shown in Figure \ref{fig:mixed_env}.
We note that our order parameter will somewhat underestimate the number of crystalline particles because it does not detect particles at the edges of crystal grains.
We computed that our order parameter misclassified particles in the fluid with a rate under 0.25$\%$.

\newpage
\section{Different Stoichoimetries and RDFs}

 We present RDFs and snapshots for our results at each stoichiometry in Figures S5-7.

\begin{figure}[htbp]
    \centering
    \includegraphics[scale=0.75]{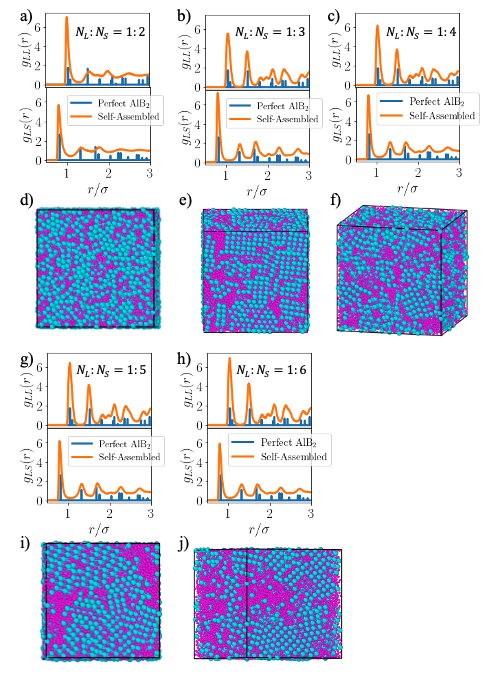}
    \caption{Radial distribution functions (RDFs) and and snapshots of NPT simulation results at (a,d) $N_L$:$N_S$ = 1:2, $P^*=70$, (b,e) $N_L$:$N_S$ = 1:3, $P^*=70$, (c,f) $N_L$:$N_S$ = 1:4, $P^*=78$, (g,i) $N_L$:$N_S$ = 1:5, $P^*=75$,   and (h,j)  $N_L$:$N_S$ = 1:6, $P^*=80$. These simulations all began in a fluid-state. The RDFs are averaged over the final 5 frames of the simulations; the snapshots  are the last frames of the simulations. We only show the RDFs for large-large and large-small interactions because that of small-small interactions tends to be dominated by fluid-like small particles. At both stoichiometries, we see crystal grains in both the snapshots and the RDFs. Visual inspection and comparing the RDFs to the perfect ones for AlB$_2$ show the crystal structure to be that of AlB$_2$.
    }
    \label{fig:alb2xs}
\end{figure}

\begin{figure}[htbp]
    \centering
    \includegraphics[scale=0.7]{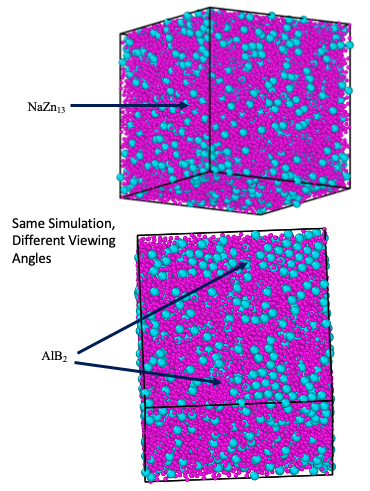}
    \caption{Snapshots showing different angles of an NPT simulation run at $N_L$:$N_S$ = 1:9 and $P^*=98$. The simulation began in a fluid-state. Both AlB$_2$ and NaZn$_{13}$ self-assemble in the simulation, as point out by the arrows. The crystal grains are small in both cases.
    }
    \label{fig:structuremix}

\end{figure}

\begin{figure}[htbp]
    \centering
    \includegraphics[scale=0.46]{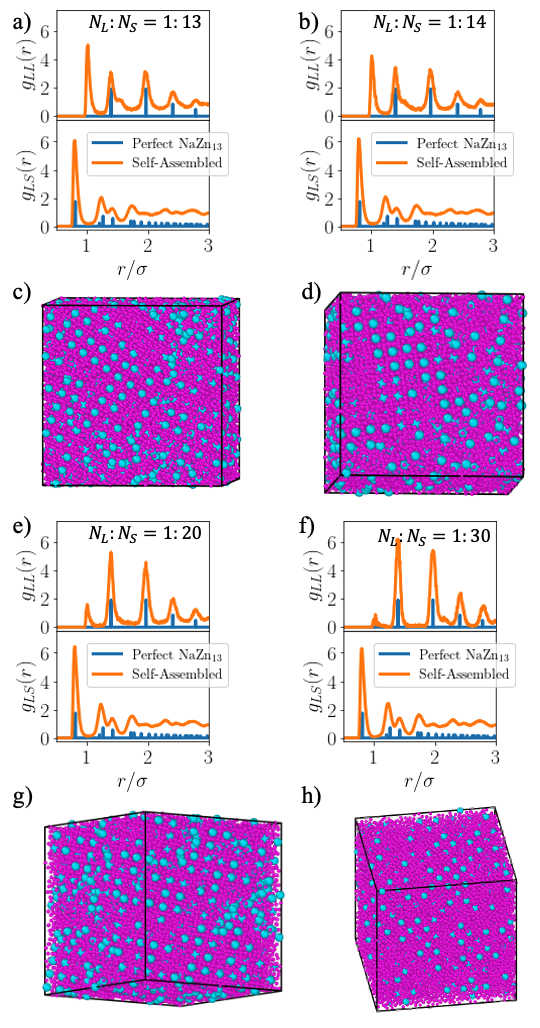}
    \caption{Radial distribution functions (RDFs) and and snapshots of NPT simulation results at (a,c) $N_L$:$N_S$ = 1:13, $P^*=98$, (b,d) $N_L$:$N_S$ = 1:14, $P^*=98$, (e,g)  $N_L$:$N_S$ = 1:20, $P^*=98$, and (f,h) $N_L$:$N_S$ = 1:30, $P^*=98$. These simulations all began in a fluid-state. The RDFs are averaged over the final 5 frames of the simulations; the snapshots  are the last frames of the simulations. We only show the RDFs for large-large and large-small interactions because that of small-small interactions tends to be dominated by fluid-like small particles. At both stoichiometries, we see crystal grains in both the snapshots and the RDFs. Visual inspection and comparing the RDFs to the perfect ones for NaZn$_13$ show the crystal structure to be that of NaZn$_13$. We note presence of an unexpected peak at $r/\sigma=1$; this is due to large particles in contact with each. There are no such contacts in perfect NaZn$_13$, but they occur in our self-assembly due to some large particles not being incorporated into the crystal.
    }
    \label{fig:nazn13xs}
\end{figure}

\newpage

\bibliography{bibliography}